\documentstyle[12pt]{article}
\newtheorem{lem}{Lemma}[section]
\newtheorem{Definition}{Definition}[section]
\newtheorem{Proposition}{Proposition}[section]
\newtheorem{Example}{Example}[section]
\newtheorem{corollary}{Corollary}[section]

\newcommand{\calA}{\mbox{${\cal A}$}}

\newtheorem{theorem}{Theorem}[section]

\begin{document}

\title{Observables in a Noncommutative Approach to the
Unification of Quanta and Gravity. A Finite Model}

\author{%
Leszek Pysiak\thanks{Department of Mathematics and Information
Science, Warsaw University of Technology, Plac Politechniki 1,
00-661 Warsaw, Poland.},
\and
Michael Heller\thanks{
Vatican Observatory, V-00120 Vatican City State. Correspondence
address: ul. Powsta\'nc\'ow Warszawy 13/94,
33-110 Tarn\'ow, Poland.  E-mail: mheller@wsd.tarnow.pl}
\and Zdzis{\l}aw Odrzyg\'o\'zd\'z\thanks{%
Department of Mathematics and Information Science, Warsaw
University of Technology Plac Politechniki 1, 00-661 Warsaw,
Poland.} \and
\and Wies{\l}aw Sasin{\thanks{%
Department of Mathematics and Information Science, Warsaw
University of Technology, Plac Politechniki 1, 00-661 Warsaw,
Poland.}}}
\date{\today}
\maketitle

\begin{abstract}
We further develop a noncommutative model unifying quantum
mechanics and general relativity proposed in {\it Gen. Rel.
Grav.\/} (2004) {\bf 36}, 111-126. Generalized symmetries of the
model are defined by a groupoid $\Gamma $ given by the action of
a finite group on a space $E$. The geometry of the model is
constructed in terms of suitable (noncommutative) algebras on
$\Gamma $. We investigate observables of the model, especially
its position and momentum obesvables. This is not a trivial
thing since the model is based on a noncommutative geometry and
has strong nonlocal properties.  We show that, in the position
representation of the model, the position observable is a
coderivation of a corresponding coalgebra, ``coparallelly'' to
the well known fact that the momentum observable is a derivation
of the algebra. We also study the momentum representation of the
model. It turns out that, in the case of the algebra of smooth,
quickly decreasing functions on $\Gamma $, the model in its
``quantum sector'' is nonlocal, i.e., there are no nontrivial
coderivations of the corresponding coalgebra, whereas in its
``gravity sector'' such coderivations do exist. They are
investigated.
\end{abstract}

\section{Introduction}
Noncommutative geometry is not only quickly developing branch of
mathematics \cite{Connes,Varilly,Landi,Madore}, but also finds
interesting applications in physics. Within its conceptual
framework several approaches have been elaborated to unify
relativity and quanta (see, for example,
\cite{Chamsed01,ChamCon96,Chamsedetal,Connes96,Madmour,MadSae}),
and noncommutative methods more and more deeply penetrate into
superstring or M-theories (first indication \cite{ConDougSchw},
first result, \cite{SeibWitt}, reviews
\cite{Karczmarek,KonSchw02}). In a series of works 
\cite{HSL,Towards,Reduction} we have proposed a scheme for
unifying general relativity and quantum mechanics based on a
noncommutative algebra related to a transformation groupoid
which plays the role of generalized symmetries of the model. In
\cite{Finite} we have tested the scheme by constructing a simplified
(but still mathematically interesting) model and computing many
of its details. In the present work, we discuss observables of
this model; in particular, position and momentum observables. It
is a common knowledge that the algebra structure of observables
is a crucial ingredient of the standard formulation of quantum
mechanics; our analysis has disclosed that its coalgebra
structure is also implicitly present in the standard approach.

The present paper is thus a continuation of \cite{Finite} (we
will refer to it as to a ``previous work''), but, for the
reader's convenience, we shortly summarize its main features and
results. We construct our transformation groupoid in the
following way. Let $\tilde{E}$ be a differential manifold and
$\tilde{G}$ a group acting smoothly and freely on $\tilde{E}$.
In consequence, we have the bundle $(\tilde{E},
\pi_M, M = \tilde{E}/\tilde{G})$. We can think of it
as of the frame bundle, with $\tilde{G}$ the Lorentz group, over
space-time $M$. Now, we choose a finite subgroup $G$ of
$\tilde{G}$, a cross section $S: M \rightarrow \tilde{E}$ of the
above bundle (it need not be continuous), and define $E =
\bigcup_{x \in M}S(x)G$.  The fact that $G$ acts freely (to the
right) on $E$, allows us to define the transformation groupoid
structure on the Cartesian product $\Gamma = E \times G$ (for
details see the previous work).

The choice of the cross section $S: M \rightarrow \tilde{E}$ can
be thought of as the choice of a gauge for our model. To be more
precise every $\gamma \in \Gamma $ can be presented in the form
$\gamma = (S(x)g, \bar g)$ where $g, \bar g \in G$. The set of
all $\gamma $'s with the beginning at $p \in E$ is denoted by
$\Gamma_p$ (a ``fiber'' of $\Gamma $ over $p$). We define the
Hilbert space $L^2(\Gamma_p) = \{u: \Gamma_p \rightarrow {\bf
C}:\Sigma_{g \in G}|u(S(x)g_0,g)|^2 < \infty \}$ with the scalar
product $\langle u,v\rangle = \Sigma_{g \in G}
\bar u(S(x)g_0,g)v(S(x)g_0,g)$. If $L_h: L^2(\Gamma_p)
\rightarrow L^2(\Gamma_p)$ is a left translation, it is
straightforward to show that $L_h$ is a unitary operator with
respect to the above scalar product. It transforms $S(x)$ into
$S(x)h$ (for each fibre independently). In this sense, our
choice of the gauge is unique up to unitary transformations.

The groupoid $\Gamma $ is a key structure of our model. It
represents a space, the elements of which are symmetry
operations of the model. The algebra ${\cal A} = C^{\infty
}(\Gamma , {\bf C})$ of smooth complex valued functions on
$\Gamma $ (if necessary, we shall assume that they vanish
at infinity) plays the role of an algebraic counterpart of this
symmetry space. In the previous work, we have reconstructed
geometry of the groupoid $\Gamma = E \times G$ (including
generalized Einstein's equations) in terms of this algebra. The
Cartesian product structure of the groupoid $\Gamma $ has
enabled us to consider its two natural components. By projecting
the full geometry into the $E$-direction we recover the usual
space-time geometry and, consequently, the standard general
relativity. It is a remarkable fact that this can equivalently
be achieved by suitably averaging elements of the algebra ${\cal
A}$ (see below Section 3). On the other hand, the regular
representation $\pi_p: {\cal A} \rightarrow {\rm End}({\cal
H}_p)$ of the groupoid algebra ${\cal A}$ in a Hilbert space
${\cal H}_p$, for $p\in E$, leads to the $G$-component of the
model which can be considered as its quantum sector.

In the present work, we develop the model by considering its
observables, especially its position and momentum observables.
This is not a trivial thing. The model is based on a
noncommutative algebra ${\cal A}$ and to determine these two
observables means to disentangle its local features (essentially
rooted in the center of the algebra ${\cal A}$) from its
nonlocal properties.

The organization of our material runs as follows.  In Section 2,
we present the preliminaries of the model, and study some
properties of the subalgebra ${\cal A}_{proj} \subset {\cal A}$
which is isomorphic to the center of ${\cal A}$, and serves to
reconstruct the standard geometry of space-time $M$. In Section
3, we formulate the eigenvalue equation for observables of the
model, and show that by averaging functions belonging to ${\cal
A}_{proj}$ we obtain functions on $M$.  The position observable
is discussed in Section 4. We show that, in the position
representation of the model, the position observable is a
coderivation of a corresponding coalgebra, whereas the momentum
observable is a derivation of the algebra (as it is well known).
This remains valid in the usual quantum mechanics (see below,
Example 1).

In Section 5, we study derivations and coderivations of the
group algebra and its dual in the case of a finite group, and,
in Section 6, we extend this analysis to the algebra and the
dual coalgebra on the groupoid $\Gamma = E \times G$. To do so
we limit ourselves to the case $M = {\bf R}^n$, and consider the
space ${\cal S}$ of smooth, quickly decreasing functions on
$\Gamma $ (the Schwarz space). It can be equipped with the
algebra structure in two ways which gives us two algebras,
${\cal S}_1$ and ${\cal S}_2$, on $\Gamma $. We then extend the
space ${\cal S}$ on $\Gamma $ to a distribution space ${\cal
S}'$ and obtain, correspondingly, two coalgebras ${\cal S}'_1$
and ${\cal S}'_2$ on $\Gamma $. We show that ${\cal S}'_2$
defines the position representation of our model, and ${\cal
S}'_1$ its momentum representation. We also demonstrate that a
coderivation of the coalgebra ${\cal S}'_2$ is the position
observable in the position representation, whereas a
coderivation of the coalgebra ${\cal S}'_1$ is the position
observable in the momentum representation. It turns out that in
the $G$-component of the model there is no localization (there
exist no non-trivial corresponding coderivations). This remains
in agreement with the ``noncommutative paradigm'' of this aspect
of the model.  In the $E$-component of the model, the
corresponding coderivations do exist and the local properties
are preserved.

Finally, in Section 7, we present the position and momentum
observables in the elegant language of a sheaf of algebras on
the groupoid $\Gamma $.

\section{Preliminaries}
Let, as usually, $\Gamma =E\times G$, with $G$ a finite group,
be a transformation groupoid. We shall consider the algebra
${\cal A}=C^{\infty}_ 0(\Gamma ,{\bf C})$ of smooth, complex
valued functions on $\Gamma$ vanishing at infinity, i.e.,
functions vanishing at infinity on every connected component of
$\Gamma $ diffeomorphic with $M$. Hermitian elements of this
algebra are candidates for being observables of the model.
However, the true observable should leave some traces in
space-time $ M$ where it could be registered by a measuring
device. This is guaranteed by the following construction.
\par
Let $C_0^{\infty }(M)$ denote the algebra of smooth functions on
$ M$ vanishing at infinity. We define
\[{\cal A}_{proj}=pr^{*}(C_0^{\infty }(M))\]
where $pr=\pi_M\circ\pi_E$.  ${\cal A}_{proj}$ is an algebra
without unit. Let $ (C_0^{\infty }(M))_M$ denote the set of
complex functions $f:M\rightarrow {\bf C}$ such that, for any
such function and for any $ x\in M$, there exists an open
neighborhood $U$ of $x$ and a function $\phi\in C_0^{\infty
}(M)$ satisfying the condition $f|U=\phi |U$.  We will say that
functions of $ (C_0^{\infty }(M))_M$ are {\em localized to $M$}.
\par
\begin{lem} \rm
$(C_0^{\infty }(M))_M=C^{\infty }(M)$ where $C^{\infty }(M)$
denotes the algebra of all smooth functions on $M$.  \end{lem}
\noindent
{\bf Proof}$\; $ The inclusion $C^{\infty }(M)\subset
(C_0^{\infty }(M))_M$ is obvious.  It is enough to show that
every continuous function $f:M\rightarrow {\bf C}$ belongs to $
(C_0^{\infty }(M))_M$. But this is indeed the case. For any $
$$x\in M$ there exists an open neighborhood $ U$ of $x$ and a
function $\phi\in C_0^{\infty }(M)$ such that $\phi |U=1$ ; of
course $\phi f\in C_0^{\infty }(M$), and $f| U=\phi f|U$. $\Box$
\par
Although the algebra ${\cal A}_{proj}$ does not contain constant
functions, we can always -- on the strength of the above lemma
-- recover them locally. Let us also notice that $({\cal
A}_{proj})_{\Gamma}$ is a subalgebra of the algebra $ ({\cal
C}_0^{\infty }(\Gamma , {\bf C}))_{\Gamma}$. Therefore, we can
safely assume that the Hermitian elements of ${\cal A}_{ proj}$
represent observables of the model.
\par
Let us now consider the regular representation $\pi_p:{\cal A}
\rightarrow {\cal B}({\cal H}_p)$ of the algebra ${\cal A}$ in
the Hilbert space ${\cal H}_p=L^2(\Gamma_p)$ given by
\[(\pi_p(a)\psi )(\gamma )=(a*\psi )(\gamma ) = \sum_{\gamma_1
\in \Gamma_p} a(\gamma \circ \gamma_1^{-1})\psi (\gamma_1)\] for
$\gamma \in \Gamma_p$. Let further $I_p:L^2(G)\rightarrow
L^2(\Gamma_p)$ be te obvious isomorphism of Hilbert spaces. For
every $a\in {\cal A}$ we define
\[\bar{\pi}_p(a)=I_p^{-1}\circ\pi_p(a)\circ I_p.\]
clearly, $\bar{\pi}_p(a)\in {\cal B}(L^2(G))$.
\par
\begin{lem} \rm
$L_{g_0}^{-1}\bar{\pi}_p(a)L_{g_0}=\bar{\pi}_{pg_0}(a)$ for
every $ g_0\in G$.
\end{lem}

\noindent
{\bf Proof} Let $\psi_p\in L^2(\Gamma_p)$ and $\psi\in 
L$$^2(G)$. We have $\psi_p=I_p(\psi )$, and we compute
\[\begin{array}{rcl}
(\pi_{pg_0(a)}\psi_{pg_0})(\bar{\gamma })&=&(a*\psi_{p
g_0})(\bar{\gamma })\\
&=&\sum_{\gamma_1\in\Gamma_{pg_0}}a(\bar{\gamma}\circ\gamma_
1^{-1})\psi_{pg_0}(\gamma_1)\\
&=&\sum_{g_1\in G}a(pg_0g_1,g_1^{-1}g)\psi (g_1)\\
&=&\sum_{g_1'\in G}a(pg_1',g^{\prime -1}_{1^{}}g_0g)\psi 
(g_0^{-1}g_1').\end{array}
\]
In the last line we have made the substitution $g_0g_1\mapsto 
g_1'$. Now, let 
$\bar{\psi }=L_{g_0}\psi$, and we compute
\[\begin{array}{rcl}
(\pi_p(a)\bar{\psi}_p)(\gamma )&=&\sum_{\gamma_1\in\Gamma_
p}a(\gamma\circ\gamma_1^{-1})\bar{\psi }(\gamma_1)\\
&=&\sum_{g_1\in G}a(pg_1,g_1^{-1}g)\bar{\psi }(p,g_1)\\
&=&\sum_{g_1\in G}a(pg_1,g_1^{-1}g)\bar{\psi }(g_1)\\
&=&\sum_{g_1\in G}a(pg_1,g_1^{-1}g_0g)\bar{\psi
}(g_1).\end{array} 
\]
In the last line we have substituted $g\mapsto g_0g$. We thus
have
\[L_{g_0}^{-1}({\rm l}\,{\rm e}{\rm f}{\rm t}\;{\rm h}{\rm a}
{\rm n}{\rm d}\;{\rm s}{\rm i}{\rm d}{\rm e})=({\rm r}{\rm i}
{\rm g}{\rm h}{\rm t}\;{\rm h}{\rm a}{\rm n}{\rm d}\;{\rm s}
{\rm i}{\rm d}{\rm e}). \; \Box \] 
\par
We now define the norm in the algebra ${\cal A}$ in the
following way
\[||a||={\rm s}{\rm u}{\rm p}_{p\in E}||\pi_p(a)||.\]
The above lemma can be written in the form
\[\pi_{s(x)g}=L_g^{-1}\circ\pi_{s(x)}\circ L_g,\]
hence
\[||\pi_{s(x)g}(a)||=||\pi_{s(x)}(a)||.\]
We see that supremum is taken over $M$, i.e.,
\[||a||={\rm s}{\rm u}{\rm p}_{x\in M}||\pi_{s(x)}(a)||.\]
For a fixed $x\in M$ we have a finally dimensional vector space
(of matrices).  In such a space all norms are equivalent. In the
matrix representation we write
\[\pi_{s(x)}(a)=(M_a(x))_{i,j},\]
and
\[||a||={\rm s}{\rm u}{\rm p}_{}(\sum_{i,j}|M_a(x)|_{i,
j}).\]
Since functions $a$ vanish at infinity, this supremum is finite.
Now, we complete the algebra ${\cal A}$ in the above norm to the
$ C^{*}$-algebra. From now on we shall always consider the
algebra ${\cal A}$ as a $ C^{*}$-algebra.
\par
A function $\psi\in L^2(\Gamma )$ is said to be $G$-{\em
invariant\/} if
\[\psi (p_1,g_1)=\psi (p_2,g_2),\]
when there exists $g\in G$ such that $p_2=p_1g$. The set of all
$ G$-invariant functions will be denoted by $L^2_G(\Gamma )$.
$L^2_G(\Gamma )$ is evidently isomorphic with $L^2(M)$. The fact
that ${\cal A}$ is a $C^*$-algebra allows us to employ in our
model the algebraic quantization method.
\par
\section{Eigenvalue Equation}
Let $a$ be a Hermitian element of ${\cal A}_{proj}$, and $
\psi\in L^2_G(\Gamma )$. The eigenvalue 
equation for the observable $a$ assumes the form
\[\pi_p(a)\psi =\lambda_p\psi\]
where $\lambda_p$ is the eigenvalue of $\pi_p(a)$. Here we have
$ p\in\pi_M^{-1}(x),\,x\in M$. For simplicity we consider a
nondegenerate case. We compute
\begin{eqnarray*}
(\pi_p(a)\psi )(\gamma )&=&(\psi *a)(\gamma )\\
&=&\sum_{g_1\in G}\psi (p,g_1)a(pg_1,g_1^{-1}g)\\
&=&\sum\tilde{\psi }(x)\tilde {a}(x)\\
&=&|G|\tilde{\psi }(x)\tilde {a}(x)\end{eqnarray*}
where $\gamma\in\Gamma_p$, and we have introduced the
abbreviations: $\tilde{\psi }(x)=\psi (p,g)$ for 
every $g\in G$, and $\tilde {a}(x)=a(p,g)$ for every $p
\in\pi_M^{-1}(x)$ and $g\in G$. If we further 
denote
\[\widetilde{(\pi_p(a)\psi )}(x)=(\pi_p(a)\psi )(\gamma ),\]
where $\gamma\in\Gamma_p$ and $p\in\pi_M^{-1}(x)$, then we
finally have
\[\widetilde{(\pi_p(a)\psi )}(x)=|G|\tilde{\psi }(x)\cdot
\tilde {a}(x).\]
Hence, the eigenvalue of this observable is
\[\lambda_p=|G|\cdot\tilde {a}(x)\]
for every $p\in\pi_M^{-1}(x).$ 
\par
As we can see, the observable $a\in {\cal A}$ indeed leaves a
trace in space-time $M$.  In the previous work we have shown
that the transition from the noncommutative geometry on the
groupoid $\Gamma$ to the classical geometry on the manifold $ M$
can be done with the help of a suitable averaging procedure of
elements of the groupoid algebra. The following lemma
establishes the equivalence of this averaging method with the
one using the subalgebra ${\cal A}_{ proj}$.
\par
\begin{lem} \rm
The averaging of a function belonging to ${\cal A}_{pro
j}$ gives a function of $C_0^{\infty }(M)$.
\end{lem}
\par
\noindent
{\bf Proof}$\;$ Let $A_a$ be the matrix representation of $a\in
{\cal A}$. Then its averaging is $\langle A_a\rangle =\frac
1{|G|}{\rm T}{\rm r}(A_a)$. If $ a\in {\cal A}_{proj}$ then
$a=pr^{*}f$ for some $f\in C_0^{\infty }(M).$ By averaging we
obtain
\[\langle A_{pr^{*}f}\rangle =\frac 1{|G|}\cdot f\cdot 
{\rm T}{\rm r}[1]=f.\]
Here [1] denotes the matrix with all entries equal to one. Of
course, ${\rm T}{\rm r}[1]=|G|$. $\Box$

\section{Position Operator as a Coderivation} 
In this section we consider the position observable of our
model. The projection $ pr:\Gamma \rightarrow M$, $pr=\pi_M\circ
\pi_E$, which is clearly connected with localization in
space-time $M$, is not a numerical function (it has no values in
{\bf R} or {\bf C}), and consequently it does not belong to the
algebra ${\cal A})$.  However, in the four-dimensional case, if
we choose a local coordinate map $x = (x^{\mu }),\,\mu =
0,1,2,3$, in $ M$ then the projection $pr$ determines four
observables in the domain ${\cal D}_x$ of $x$
\[pr_{\mu } = x^{\mu } \circ pr \]
We thus have the system of four position observables
$pr=(pr_0,pr_1,pr_2,pr_3)$. For a fixed $\mu $ one has
$pr_{\mu}\in {\cal A}_{proj}|_{{\cal D}_x}$, and it is
Hermitian.

Let us notice that  the projection $ pr:\Gamma \rightarrow M$
contains, in a sense, the information about all possible local
observables $ pr_{\mu}$. This can be regarded as a
``noncommutative formulation'' of the fact that there is no
absolute position but only the position with respect to a local
coordinate system.

From the previous work (Section 7) it follows that in the matrix
representation of the algebra ${\cal A}$ one has
\[
(\pi_q(pr_1))(\xi ) = \xi^T \cdot M_{pr_1} = x^1 \cdot \xi ,
\]
$x \in M$, in the local map, where $\xi^T$ is $\xi \in {\bf
C}^n$ transposed, and $M_{pr_1}$ denotes the matrix
corresponding to the projection $pr_1$. We see that the position
observable in the ''quantum sector'' of our model has the same
form as in the ordinary quantum mechanics. This indicates that
we are working in the position representation of the model.

We shall now demonstrate the connection between the position
operator and the coderivation of a coalgebra. Let $({\cal
A},+,m,)$ be an associative (not necessarily commutative)
algebra over the field of complex numbers ${\bf C}$.  Here $m:
{\cal A} \otimes {\cal A} \rightarrow {\cal A}$ is a product
map.  We define the dual space ${\cal A}^{*}={\rm H}{\rm o} {\rm
m}({\cal A},{\bf C})$ which has the structure of a coalgebra
with the coproduct $\Delta :{\cal A}^{*}\rightarrow {\cal
A}^{*}\otimes {\cal A}^{*}$ given by
\[\Delta (\varphi )(f,g)=\varphi (m(f\otimes g)),\]
$g,f\in {\cal A},\,\varphi\in C^{*}$. We assume that $({\cal
A} \otimes {\cal A})^* \subset {\cal A}^* \otimes {\cal A}^*$.
This is always true for finite dimensional algebras. Rather than
considering a completion of the tensor product ${\cal A}^*
\otimes {\cal A}^*$ (see \cite[Chapters 5-6]{CannasWeinstein}),
we shall slightly modify the coproduct definition (in Section 6).
\par
We recall that a {\em derivation} $X$ of the algebra ${\cal A}$
is, by definition, a linear map $ X:{\cal A}\rightarrow {\cal
A}$ satisfying the Leibniz rule
\[X\circ m=m\circ (X\otimes {\rm i}{\rm d}_{{\cal A}})+m
\circ ({\rm i}{\rm d}_{{\cal A}}\otimes X).\]
The set Der${\cal A}$ of all derivations of the algebra ${\cal
A}$ is a ${\bf {\cal A}}$-module.
\par
Let  $X^{*}:{\cal A}^{*}\rightarrow {\cal A}^{*}$ be a linear
mapping satisfying the following condition
\[\Delta\circ X^{*}=(X^{*}\otimes {\rm i}{\rm d}_{{{\cal A}}^{
*}})\circ\Delta +({\rm i}{\rm d}_{{{\cal A}}^{*}}\otimes X^{
*})\circ\Delta ,\]
which we shall call the {\em co-Leibniz\/} rule. For $\varphi
\in {\cal A}^{*}$ it can be written in the 
form
\[\Delta (X^{*}(\varphi ))=X^{*}(\varphi_{(1)})\otimes\varphi_{
(2)}+\varphi_{(1)}\otimes X^{*}(\varphi_{(2)}).\]
\begin{Definition} \rm
Let $ $${\cal A}$ be an associative not necessarily commutative
algebra, and $ {\cal A}^{*}$ its dual coalgebra. A {\em
coderivation\/} of the coalgebra ${\cal A}^{ *}$ is a linear map
$X^{*}:{\cal A}^{*}\rightarrow {\cal A}^{ *}$ satisfying the
co-Leibniz rule. \end{Definition}
\par
\begin{Proposition} \rm
Let $X:{\cal A}\rightarrow {\cal A}$ be an endomorphism of an
algebra ${\cal A}$ as a
linear space, and $X^{*}:{\cal A}^{*}\rightarrow {\cal A}^{ *}$
its adjoint endomorphism, i.e.,
\[(X^{*}(\varphi ))(a)=\varphi (X(a)),\]
$\varphi\in {\cal A}^{*},\,a\in {\cal A}.$ $X$ satisfies the
Leibniz rule if and only if $ X^{*}$ satisfies the co-Leibniz
rule.
\end{Proposition}

\noindent
{\bf Proof} Let us suppose that $X$ satisfies the Leibniz rule,
then 
\begin{eqnarray*}
(\Delta (X^{*}\varphi ))(a_1,a_2)&=&(X^{*}\varphi )(a_1
\cdot a_2)\\
&=&\varphi (X(a_1\cdot a_2))\\
&=&\varphi (X(a_1)a_2)+\varphi (a_1X(a_2))\\
&=&\Delta\varphi (X(a_1),a_2)+\Delta\varphi (a_1,X(a_2)
)\\
&=&(((X^{*}\otimes\rm id_{{\cal A}^{*}})\circ\Delta +(\rm 
id_{{\cal A}^{*}}\otimes X^{*})\circ \Delta )(\varphi ))(a_1,a_
2{}_{}),\end{eqnarray*}
and similarly in the other direction. $\Box$
\par
\begin{Example}
{\rm
Let us consider the algebra ${\cal A} = (L^1({\bf R}),*)$, where
$*$ denotes the convolution. The Fourier transformation of $f
\in L^1({\bf R})$ gives
\[ \hat f(x) = \int_{{\bf R}}e^{-ipx}f(p)dp .\]
We have
\[\int_{{\bf R}}e^{-ipx}\frac d{dp}f(p)dp = ix\int_{{\bf R}}
e^{-ipx}f(x)dx\]
or, in the abbreviated form,
\[\widehat{f^{'}}(x)=ix \cdot \hat {f}(x),\]
$x \in {\bf R}$.
\par
The dual of ${\cal A}=L^1({\bf R})$ is ${\cal A}^{*}=L^{
\infty}({\bf R})$ (with the pointwise multiplication). Let us
denote  
\[(X^{*}\varphi )(x)= x\varphi (x)\]
for $\varphi \in L^{\infty }({bf R})$. $X^*$ is an operator
adjoint to the operator $X: {\cal A} \rightarrow {\cal A}$ given
by $X = -i\frac {d}{dp}$.
\par
Since {\bf R} is an Abelian group, as a coproduct in $L^{
\infty}({\bf R})$, we have
\[\Delta\varphi (x_1,x_2)=\varphi (x_1+x_2)\]
for $\varphi\in L^{\infty}({\bf R})$, $x_1,x_2\in {\bf R}$.  It
can be readily checked that $ X_1^{*}$ is a coderivation of the
coalgebra $L^{\infty}({\bf R})$. This coderivation is called the
{\it position operator in the position representation\/}.}
\end{Example}
\par\ 
The above example suggests that in noncommutative
generalizations of quantum mechanics we can treat coderivations
of suitable coalgebras as counterparts of the position operator
(in the position representation).  We will explore this
possibility in the following sections.

\section{Derivations and Coderivations on a Finite Group}
In this section, we apply considerations of the previous section
to the quantum mechanics on a finite group $G$. In this case, we
have the group algebra $H={\bf C}G$ of formal linear
combinations of elements of $G$ (which corresponds to the
momentum representation) and the algebra $H^{*}={\bf C}(G)$ of
complex functions on $G$ with pointwise multiplication
(corresponding to the position representation of the model). Let
us notice that both $H$ and $H^*$ are bialgebras with coproducts
being conjugations of products of the corresponding dual
algebras. We assume that $G$ is a non-Abelian group.
\par
Let us consider the Fourier transformation of a finite group $G$
\[{\cal F}:G\rightarrow\Pi M_{d_{\lambda}}({\bf C})\equiv 
{\cal M},\]
here $d_{\lambda}$ is the dimension of the representation from
the class $
\lambda$, $\lambda \in \hat{G}$ where $\hat{G}$ is the dual
object of $G$, given by
\[{\cal F}(g)=(T_{\lambda}(g))_{g\in G}\]
with $T_{\lambda}(g)\in M_{d_{\lambda}}({\bf C})$. This is
extended by linearity to the whole of the group algebra ${\cal
F}:{\bf C}G\rightarrow {\cal M}$ \cite[Chapter 12]{Kirillov}. As
it is well known, $ {\cal F}$ is the isomorphism of algebras. We
thus can change the algebra ${\bf C}G$ into the corresponding
matrix algebra. The latter algebra has only inner derivations
with dim(Inn$(M_n({\bf C}))=n^2-1$.
\par
\begin{lem} \rm
Let ${\cal A}$ be a unital algebra, $J$ an ideal defined by a
central idempotent, i.e., $J=e{\cal A}={\cal A}e,\,e^2=e,\,e\in
{\cal Z}({\cal A} )$, and $D$ a derivation of the algebra ${\cal
A}$. Then
\[DJ\subset J,\;\;\;De=0.\]
\label{LemA} \end{lem}
\par
\noindent
{\bf Proof} $\;$ The fact that $e$ is idempotent implies
\[De=D(e^2)=(2De)e\in J.\]
From $1=e+e',$ where $e'=1-e$ is also an idempotent, it follows
that $ee'=e'e=0$, and we have
\[D1=De+De'=0.\]
Therefore,
\[De=-De',\]
and
\[D(ae)=(Da)e\in J\]
for any $a\in {\cal A}$. Thus $DJ\subset J.$$\Box$
\par
\begin{lem} \rm
Any isomorphism of algebras determines the isomorphism of the
corresponding derivation spaces in such a way that the inner
derivations are transformed into the inner derivations. $\Box$
\label{LemB} \end{lem}
\par
\begin{theorem} \rm
In the algebra ${\bf C}G$ there are only inner derivations, and
in the algebra $ {\bf C}(G)$ there are no nonzero derivations.
\end{theorem}
\par
\noindent
{\bf Proof} (1) Let $J_{\lambda_0}\subset {\cal M}$ be an ideal
of $ {\bf C}G$ such that
\[J_{\lambda_0}=\{(0,0,\ldots ,A_{\lambda_0},0,\ldots 0
):A_{\lambda_0}\in M_{\lambda_0}\}.\] 
If $\bar {D}$ is a derivation of the algebra ${\cal M}$ then, on
the strength of Lemma \ref{LemA}, $\bar {D}J_{\lambda_0}\subset
J_{\lambda_0}$.  Therefore $ J_{\lambda_0}$ is isomorphic with
$M_{d_0}({\bf C})$ which has only inner derivations. Hence
\[\bar {D}|_{J_{\lambda_0}}={\rm a}{\rm d}B_{\lambda_0}\]
where $B_{\lambda_0}\in J_{\lambda_0}$.
\par
Let now $B=(B_{\lambda})_{\lambda\in\hat {G}}$ be the sequence
as above. Therefore, $
\bar {D}={\rm a}{\rm d}B$ and, on 
the strength of Lemma \ref{LemB}, in the algebra ${\bf C}
G$  there exist only inner derivations. 
\par
(2) The element $\delta_{g_0}$ is central idempotent in $
{\bf C}(G)$, i.e., $\delta_{g_0}\cdot\delta$$_{g_0}=\delta_{
g_0}$, 
and such elements form a basis $\{\delta_g\}$ in ${\bf C}
(G)$. Therefore, $D(a)=0$ for every 
$a\in {\bf C}(G)$ and any derivation $D$ of ${\bf C}(G)$. $
\Box$
\par
To sum up, in the bialgebra ${\bf C}G$ there exist only
(nonzero) derivations, and in the bialgebra ${\bf C}(G)$ only
(nonzero) coderivations.
\par

Our goal is now to discuss the position observable on a finite
group $G$. It is given by a coderivation of the coalgebra ${\bf
C}(G)$, and could be found as an adjoint of the derivation of
the algebra ${\bf C}G$. But the latter algebra has only nonzero
inner derivations; they are of the form
\[X_{g_0}(g)=({\rm a}{\rm d}g_0)(g)=g_0g-gg_0\]
with $g_0\in G$. Therefore, if $f\in {\bf C}(G)$, we dually have
\[(X^{*}_{g_0}(f))(g)=f(X_{g_0}g)=f(g_0g)-f(gg_0).\]
The eigenvalue equation assumes the form
\[(X^{*}_{g_0})f=\lambda\cdot f\]
where $\lambda\in {\bf C}$, or
\[f(g_0g)-f(gg_0)=\lambda\cdot f(g).\]
\par
This equation has non-trivial solutions. For instance, it can be
easily seen that the eigenspace corresponding to the eigenvalue
$\lambda =0$ is the space of central functions.\footnote{We
remind that a function $f$ is central if $f(g)=f(g_0gg_0^{-1})$
for any $ g,\,g_0\in G$.} Therefore, we can have a well
determined localization on a finite group. This remains in
agreement with the fact that $ {\bf C}(G)$ is a commutative
algebra.
\par 
\section{Localization on a Trans\-for\-mation Grou\-poid} 
In this section, we extend the above analysis to the case of the
transformation groupoid $\Gamma =E\times G$ where $G$ is a
finite (non-Abelian) group. To do this we limit our
considerations to the case when the base space $M={\bf R}^n$.
We extend the space of functions on $\Gamma$ to the distribution
space on $\Gamma $.  Let then ${\cal S}={\cal S}(\Gamma ,{\bf
C})$ be the space of smooth, quickly decreasing functions on $
\Gamma$, called also the {\it Schwarz space\/} (we recall that
the Schwarz space on $ {\bf R}^n$ is the vector space ${\cal S}$
of smooth functions on ${\bf R}^ n$ such that for every $\phi\in
{\cal S}$, $\phi$ and its derivatives decrease more rapidly than
any power of $ 1/|x|$, $x\in {\bf R}^n$, when $|x|$ goes to
infinity, \cite[p. 474]{3P}). In the following we consequently
use the matrix representation of ${\cal S}$. We thus have
\[{\cal S}={\cal S}({\bf R}^n)\otimes M_n({\bf C}).\]
A (nonunital) algebra structure in ${\cal S}$ can be introduced
in two ways: 
\begin{enumerate}
\item 
\[m_1[(f\otimes A)\otimes (g\otimes B)]=fg\otimes A\cdot 
B,\]
$f,g\in {\cal S}({\bf R}^n),\,A,B\in M_n({\bf C})$, where $ f$
and $g$ are multiplied pontwise, and $A$ and $B$ are multiplied
in the usual matrix way; we will write ${\cal S}_1=({\cal S}
,m_1)$.
\item
\[m_2[(f\otimes A)\otimes (g\otimes B)]=f*g\otimes A\cdot 
B\]
where $*$ denotes the usual convolution of functions; in this
case we write ${\cal S}_2=({\cal S},m_2)$.
\end{enumerate}

We should distinguish two kinds of derivations of the above
algebras: 
\begin{enumerate}
\item
{\em Vertical derivations}, $X_A,\,A\in M_n({\bf C}
)$, are of the form
\[X_A={\rm i}{\rm d}\otimes {\rm a}{\rm d}A,\]
or, for $ \in {\cal S}({\bf R}^n)$,
\[X_A(f\otimes B) = [1\otimes A, f\otimes B]\]
on the basis elements and extended by linearity.
\item
{\em Horizontal derivations\/} are linearly generated by
\[X_k=D_k\otimes {\rm i}{\rm d}_{M_n({\bf C})}\]
where $k=0,1,\ldots , n$ and $D_k=\frac 1i\frac
{\partial}{\partial x_k}$, i.e.,
\[X_k(g\otimes B)=\frac 1i\frac {\partial g}{\partial x_
k}\otimes B,\]
and are extended by linearity.
\end{enumerate}

As a distribution space we assume
\[{\cal S}'={\cal S}'({\bf R}^n)\otimes M_n({\bf C})\]
where ${\cal S}'({\bf R}^n)$ is the dual of ${\cal S}({\bf R}^
n)$ (but ${\cal S}'$ is not the dual of ${\cal S}$). The space
${\cal S}'$ has no algebra structure. However, if we slightly
modify the usual condition for coproduct, we can introduce in it
(in two ways) the coalgebra structure. The usual coproduct would
be the mapping $\Delta : S' ({\bf R}^n)\rightarrow {\cal
S}'({\bf R}^n)\otimes {\cal S}'({\bf R}^n)$, whereas we assume
$\Delta_i : {\cal S}'({\bf R}^n)\rightarrow ({\cal S}^{}( {\bf
R}^n)\otimes {\cal S}({\bf R}^n))'$, $i=1,2$, (which is also
valid in the usual approach for finally dimensional cases), and
for $T\in {\cal S}'({\bf R}^n)$ we define
\[(\Delta_iT)(f\otimes g)=T(m_i(f\otimes g))\]
where $m_i$ are restricted to the first factors in the
corresponding tensor products, or $\Delta_i=m_i^{*}$, i.e., our
coproduct is a dual 
homomorphism of linear spaces. 
This shows that our definition is very natural. However, we must 
accordingly adapt the associativity condition. We introduce the 
following notation 
\[\Delta_i\bar{\otimes }{\rm i}{\rm d}:({\cal S}({\bf R}^
n)\otimes {\cal S}({\bf R}^n))'\rightarrow ({\cal S}({\bf R}^
n)\otimes {\cal S}({\bf R}^n)\otimes {\cal S}({\bf R}^n
))'\]
where $\Delta_i\bar{\otimes }{\rm i}{\rm d}=(m_i\otimes 
{\rm i}{\rm d})^{*}$ and, analogously, ${\rm i}{\rm d}\bar{
\otimes}\Delta_i=({\rm i}{\rm d}\otimes m_i)^{*}$. With this 
notation our associativity condition reads
\[(\Delta_i\bar{\otimes }{\rm i}{\rm d})\circ\Delta_i=(
{\rm i}{\rm d}\bar{\otimes}\Delta_i)\circ\Delta_i.\]
It can be easily checked that the above defined coproducts
satisfy this condition. 

We define the coproduct on $M_n({\bf C})$ in the usual way; in
the basis $\{E_{ij}\}$ we have
\[\Delta_{M_n({\bf C})}=E_{ij}\otimes E_{ij},\]
and we extend this by linearity.

Finally, we set
\[\bar{\Delta}_i=\Delta_i\otimes\Delta_{M_n({\bf C})}.\]
As in Section 4, coderivations of ${\cal S}'$ are defined to be
linear mappings of ${\cal S}'$ adjoint to the derivations of
${\cal S}$. With the coproduct defined as above, this requires a
slight modification of the co-Leibniz rule. If $ X$ is a
derivation of ${\cal S}({\bf R}^n)$ then $X^{*}$ is a
coderivation of $ {\cal S}'({\bf R}^n)$, i.e., it satisfies
the co-Leibniz rule
\[\Delta\circ X^{*}=(X^{*}\bar{\otimes }{\rm i}{\rm d})
+({\rm i}{\rm d}\bar{\otimes }X^{*})\]
where we define
\[X^{*}\bar{\otimes }{\rm i}{\rm d}=(X\otimes {\rm i}{\rm d}
)^{*},\]
and analogously for ${\rm i}{\rm d}\bar{\otimes }X^{*}$.
Remembering that the distributional derivative of $T$ is defined
to be
\[(D_kT)(f)=-T(D_kf),\]
we have $X=D_k:{\cal S}_1({\bf R}^n)\rightarrow {\cal S}_1( {\bf
R}^n)$ with $D_k=\frac 1i\frac {\partial}{\partial x_k}$, and
$X^{*}:{\cal S}'_1({\bf R}^n)\rightarrow {\cal S}'_1 ({\bf R}^n)$
with $X^{*}=-D_k$. It can be readily checked that $-D_k$
satisfies the above modified co-Leibniz rule.

Now, we consider the distributional Fourier transform $
{\cal F}^{*}:{\cal S}_1'({\bf R}^n)\rightarrow {\cal S}_
2'({\bf R}^n)$ which is given by
\[({\cal F}^{*}T)(f)=T({\cal F}f)\]
where ${\cal F}f$ is to be understood as
\[({\cal F}f)(x)=\int_{{\bf R}^n}f(t)e^{-itx}d\mu_k(x)\]
with $d\mu_k(x)=dx/(2\pi )^{n/2}$, normalized Lebesgue measure. $
{\cal F}$ is a continuous algebra isomorphism of ${\cal
S}_1({\bf R}^n)$ into $ {\cal S}_2({\bf R}^n)$, and ${\cal
F}^{*}$ is a continuous linear isomorphism of ${\cal S}_1'({\bf
R}^n)$ into ${\cal S}_2'({\bf R}^n)$ which is additionally a
coalgebra map, i.e., it satisfies the condition
\[\Delta_2\circ {\cal F}^{*}=({\cal F}^{*}\bar{\otimes }
{\cal F}^{*})\circ\Delta_1.\]

One can check that, for $X^{*}=-D_k$, one has
\[{\cal F}^{*}(X^{*}T)=x_k({\cal F}^{*}T),\]
or, for $\hat {T}\in {\cal S}'_2({\bf R}^n)$, 
\[\tilde {X}^{*} \hat {T}=x_k\cdot\hat {T}.\]
It can be also verified that $\tilde {X}^{*}
\hat {T}$ is a coderivation of ${\cal S}'_2({\bf R}^n)$.

The question remains to be answered of whether do exist nonzero
vertical derivations, i.e., the derivations of the form
\[\tilde {Y}^{*}={\rm i}{\rm d}\otimes Y^{*}\]
where $Y^{*}$ is a coderivation of the coalgebra $M_n({\bf C}
)$? Clearly, the answer is ``no'' since there do not exist
nonzero derivations of $M_n({\bf C} )^{*}$ (with the pointwise
multiplication).

Now, we should collect the results and conclusions of this
lengthy analysis.  First, we make this clear that by the
position representation of our model we understand the
representation in which the position operator has the form of
the multiplication by a coordinate, and by the momentum
representation the one in which the momentum operator has the
form of the multiplication by a coordinate. In this sense, the
coalgebra $ {\cal S}'_2$ defines the position representation of
our model, and the coalgebra ${\cal S}'_1$ its momentum
representation, and we have

\begin{corollary} \rm
\begin{enumerate}
\item
The operator $\tilde {X}_k^{*}\otimes {\rm i}{\rm d}_{M_n({\bf
C} )}$ is a coderivation of the coalgebra ${\cal S}'_2$, and it
is the position observable in the position representation.
\item
The operator $X^{*}\otimes {\rm i}{\rm d}_{M_n({\bf C})}$ is a
coderivation of the coalgebra ${\cal S}'_1$ , and it is the
position observable in the momentum representation. $\Box $
\end{enumerate}
\end{corollary}

We can see that the localization on the groupoid $\Gamma $ comes
only from the ``horizontal component'' of our model which
reflects essentially the space-time geometry; whereas its
``vertical component'', representing the quantum sector of the
model, is entirely nonlocal.

\section{A Sheaf Structure on the Groupoid}
In this section, we show how could one elegantly describe, by
exploring a sheaf structure on the transformation groupoid
$\Gamma $, the position and momentum observables of our model.
On the Cartesian product $\Gamma=E\times G$ there exists the
natural product topology; however, we shall consider a weaker
topology in which the open sets are of the form $\pi_E^{- 1}(U)$
where $U$ is open in the manifold topology $\tau_E$ on $E$.
Every such open set is also open in the topology
$\tau_E\times\tau_G$. Indeed, every such set is given by
$\pi_E^{-1}(U)=U\times G$.

Let $\underline {{\cal A}}$ be a functor which associates with
an open set $ U\times G$ the involutive noncommutative algebra
${\cal A} (U\times G)$ of smooth compactly supported complex
valued functions with the ordinary addition and the convolution
multiplication.  As it can be easily seen, $\underline { {\cal
A}}$ is a sheaf of noncommutative algebras on the topological
space $(\Gamma ,\pi_E^{-1} (\tau_E))$.

The projection $pr\!\!:\Gamma \rightarrow M$ can be {\em
locally\/} interpreted as a set of (local) cross sections of the
sheaf $\underline { {\cal A}}$ (i.e.  as a set of position
observables).  Indeed, for the domain ${\cal D}_x$ of any
coordinate map $ x=(x^0,x^1,x^2,x^3)$, the composition $x\circ
pr=(x^0\circ pr,\,x^1\circ pr,\,x^2\circ pr,\,x^3\circ pr )$ is
a set of such local cross sections of $\underline {{\cal A}}$ on
the open set $\pi_E^{-1}({\cal D}_x\times G).$ The global
mapping $ pr:\Gamma \rightarrow M$ is not a cross section of
$\underline { {\cal A}}$.

Let us notice that to a measurement result which is not a number
but a set of numbers there does not correspond a single
observable but rather a set of observables, i.  e.,  a set of
(local) cross sections of the sheaf $\underline { {\cal A}}$.

Now, we define the {\em derivation morphism\/} of the sheaf
$\underline { {\cal A}}$ over an open set
$U\in\pi_E^{-1}(\tau_E)$ as a family of mappings $
X=(X_W)_{W\subset U}$ such that
$X_W:\underline {\calA }(W)\rightarrow\underline {
{\cal A}}(W)$
is a derivation of the algebra $\underline {
{\cal A}}(W),$ and for any $W_
1,W_2$ open and 
$W_1\subset W_2\subset U$, the following diagram commutes
\vspace{1.5cm}

\hspace{-3cm} 
\unitlength=1.00mm
\special{em:linewidth 0.4pt}
\linethickness{0.4pt}
\begin{picture}(129.00,87.00)
\put(58.00,82.00){\makebox(0,0)[cc]{$\underline{\cal A}(W_2)$}}
\put(123.00,82.00){\makebox(0,0)[cc]{$\underline{\cal A}(W_2)$}}
\put(123.00,47.00){\makebox(0,0)[cc]{$\underline{\cal A}(W_1)$}}
\put(58.00,47.00){\makebox(0,0)[cc]{$\underline{\cal A}(W_1)$}}
\put(64.00,82.00){\vector(1,0){53.00}}
\put(117.00,82.00){\vector(0,0){0.00}}
\put(64.00,47.00){\vector(1,0){53.00}}
\put(58.00,76.00){\vector(0,-1){23.00}}
\put(123.00,76.00){\vector(0,-1){23.00}}
\put(90.00,87.00){\makebox(0,0)[cc]{$X(W_2)$}}
\put(90.00,51.00){\makebox(0,0)[cc]{$X(W_1)$}}
\put(52.00,65.00){\makebox(0,0)[cc]{$\rho^{W_2}_{W_1}$}}
\put(129.00,65.00){\makebox(0,0)[cc]{$\rho^{W_2}_{W_1}$}}
\end{picture}

\vspace{-3cm}

\noindent
where $\rho^{W_1}_{W_2}$ is the known restriction homomorphism.
The family of all derivation morphisms indexed by open sets
is a sheaf of ${\cal Z}({\cal A})$-modules where $ {\cal
Z}({\cal A})$ denotes the sheaf of centers of the algebras
$\underline { {\cal A}}(U),\,$$U\in\pi_E^{-1} (\tau_e)$.

Components of the momentum observable $
\bar{\partial}_{\mu}$ are cross sections of the sheaf 
of ${\cal Z}({\cal A})$-modules of derivations of the sheaf
$\underline {{\cal A}}$ over domains of coordinate maps, and the
representation $\pi_U:\underline {{\cal A}}(U)
\rightarrow\pi_U(\underline {{\cal A}}
(U))$, where $U\in\pi_E^{-1}(\tau_E)$, transfers the sheaf
structure from the groupoid $\Gamma $ to the family of operator
algebras over the topological space $(\Gamma ,\,$$\pi_
E^{-1}(\tau_E)$).

\end{document}